\documentstyle[aps,prl,multicol,epsfig]{revtex}

\newcommand {\be}{\begin{equation}}
\newcommand {\ee}{\end{equation}}

\begin{document}

\title{Spreading and localization of wavepackets in 
disordered wires in a magnetic field}
\author{Matthias~Weiss, Tsampikos~Kottos, and Theo~Geisel\\
Max-Planck-Institut f\"ur Str\"omungsforschung und Fakult\"at Physik
der Universit\"at G\"ottingen,\\ Bunsenstra\ss e 10, D-37073 G\"ottingen, Germany}
\maketitle

\begin{abstract}
We study the diffusive and localization properties of wavepackets in disordered 
wires in a magnetic field. In contrast to a recent supersymmetry approach our
numerical results show that the decay rate of the steady state changes {\em smoothly} 
at the crossover from preserved to broken time-reversal symmetry. Scaling and 
fluctuation properties are also analyzed and a formula, which was derived analytically
only in the pure symmetry cases is shown to describe also the steady state 
wavefunction at the crossover regime. Finally, we present a scaling for the variance 
of the packet which shows again a smooth transition due to the magnetic field.
\end{abstract}

\begin{multicols}{2}

In recent years, considerable progress has been made in understanding the structure
of eigenstates of quasi-one-dimensional (quasi-1D) disordered systems (see~\cite{FM94}
and references therein). Most of the investigations have been concerned with the
cases of completely preserved and totally broken time-reversal symmetry. The transition
between these limits, however, has been studied much less and little is known about 
it. The only results for the transition were based on a heuristic approach by 
Bouchaud~\cite{B91}, a semiclassical analysis by Imry and Lerner~\cite{LI95} and 
numerical studies based on transfer matrices by Pichard et al.~\cite{PSSD90}. The main 
prediction of the above studies is the doubling of the localization length in the 
limit of strong magnetic fields while a smooth transition towards this asymptotic 
limit was assumed. Similar results were obtained for quantum chaotic models~\cite{SB92}. 
Recently, the doubling of the localization length was 
observed in sub-micron thin wires of doped GaAs~\cite{KGB98}, where for increasing 
magnetic field strengths a continously decreasing activation energy was reported, 
which saturated indeed at half of its field free value. Motivated by this experiment, 
the first attempt to study such a transition using the supersymmetry technique was 
undertaken in~\cite{KE99,KE00}. These authors predict that the decay of wavefunctions 
in disordered wires in weak magnetic fields, is characterized by two localization 
lengths; the far tails decay with the length $l_{\rm tails}$ characteristic for 
completely broken time-reversal symmetry, while at shorter distances the decay 
length is $l_{\rm orig} =l_{\rm tails}/2$. 

The predictions of~\cite{KE99,KE00} imply two different temperature regimes in the 
hopping conductivity separated by a boundary which depends on the magnetic field 
strength. A direct numerical test of the above prediction by means of the transmission 
approach was carried out recently in~\cite{SB99}. The results are contradicting 
the two-scale behaviour of the wavefunction in the crossover regime. The following 
possibilities were proposed in order to explain this discrepancy~\cite{KE99,KE00,SB99}:
(a) the Borland conjecture (which connects the asymptotic decay length of the
transmittance to the asymptotic decay length of the wavefunction) breaks down
in the crossover regime between the preserved and broken time reversal symmetry,
and (b) the two-scale localization phenomenon is due to anomalously localized 
states that are irrelevant for a typical wire. The latter case appears more likely
since in~\cite{KE99,KE00} the average wavefunction was calculated whereas, in~\cite{SB99} 
under the assumption of the Borland conjecture the authors calculated the average 
of the logarithm of the wavefunction. We are going to show that none of 
the above reasons can resolve the contradictions!

In this Letter the problem of the crossover behaviour, from preserved to broken 
time reversal symmetry, is addressed from a totally different perspective. In 
particular we study the evolution of initially $\delta$-like wavepackets in a 
quasi-1D geometry under the influence of a magnetic field. We then directly 
calculate the logarithmic and the arithmetic average of the steady state 
wavefunction. We show that the extracted decay rates change smoothly as a function 
of the magnetic field for both observables. Moreover, we study the whole steady 
state distribution and its fluctuations as a function of the magnetic field. 
From our numerical data we extract an analytical formula for the asymptotic 
wavefunction in the crossover regime. Finally, we present arguments according 
to which the diffusive constant scales smoothly with respect to the magnetic 
field. This imposes a smooth scaling behaviour of the variance of the packet. 
Our predictions are confirmed by extensive numerical calculations.

The mathematical model we consider is the time-dependent Schr\"odinger 
equation on a 1D lattice,  
\be\label{eqmo}
i\,{\frac{dc_n(t)}{{dt}}}=\,\sum_{m=n-b}^{n+b}H_{nm}c_m\quad , 
\ee
where $c_n(t)$ is the probability amplitude for an electron to be at site~$n$
and $H_{nm}=H_{nm}^0 + i\alpha A_{nm}$ is a complex Hermitian Band Random 
Matrix (BRM), which is decomposed into a real symmetric matrix~$H^0$ and a real 
antisymmetric matrix~$A$ with imaginary weight $i\alpha$. The entries of the
two matrices are independent Gaussian random numbers with variance $\sigma^2=1
+\delta_{nm}$ (where $\delta_{nm}$ is the Kronecker symbol) if $|n-m|\le b$ 
and zero otherwise. The parameter~$b$ defines the hopping range between 
neighbouring sites, or, in the quasi-1D interpretation, the number of transverse 
channels along a thin wire. To relate the parameter~$\alpha$ to the magnetic 
field~$B$, we note that the perturbation of the levels is proportional to 
the magnetic flux $\Phi =k B b l_{\infty} $ through an area limited by the 
localization length. Here $k$ is a dimensionless constant of order unity 
that depends on the specific geometry of the disordered wire~\cite{KE00,E97}. 
Hence we expect $\alpha \sim \Phi/\Phi_0$, where $\Phi_0=h/e$ is the 
elementary flux quantum. 

We have integrated Eq.~(\ref{eqmo}) numerically using a Cayley scheme in 
order to preserve the norm~\cite{PRT00}. Moreover a self-expanding algorithm
was implemented to eliminate finite-size effects. Whenever the probability
of finding the particle at the edges of the lattice exceeded $10^{-15}$,
$20b$ new sites were added to each end. Since all eigenstates of the 
Hamiltonian~$H_{nm}$ are known to be exponentially localized with a localization 
length $l_{\infty}(E)\sim b^2$~\cite{loc}, the evolution of the wavepacket 
is expected to exhibit a relaxation to a steady-state distribution in the 
limit $t\rightarrow \infty$. On the basis of numerical calculations it was 
shown in~\cite{IKPRT96} that for the time reversal symmetry the asymptotic 
profile $f_s(n)\equiv|c_n(t\to \infty)|^2$ is given by the following expression:
\be\label{Gogolin}
f_s(x)=\frac{\pi ^2}{16 l_{\infty}}\int_0^\infty \frac{\eta {\rm sh}(\pi
\eta )(1+\eta ^2)^2}{(1+{\rm ch}(\pi \eta ))^2}e^{
 -{1+\eta^2 \over 4 l_{\infty}} |x|} d\eta  \;,
\ee
where~$l_{\infty}$ is the averaged (over energy) localization length. 
Equation~(\ref{Gogolin}) was later proved analytically, for quasi-1D systems 
with preserved and broken time reversal symmetry~\cite{Z97}. It is interesting to 
note that Eq.~(\ref{Gogolin}) was derived first for continous 1D models 
with white noise potential~\cite{G76} with~$l_{\infty}$ being the mean free path. 
Remarkably, in spite of the relevant difference between the 1D and quasi-1D case 
(the mean free path is of the order of~$b$ in the latter and thus much smaller 
than the localization length $l_\infty \sim b^2$), the asymptotic shape remains 
the same in both cases. From~(\ref{Gogolin}), one finds that close to the origin,
\be
f_s(x)\sim \exp (-|x|/l_{\infty});\,\,\, |x| \leq l_{\infty}, 
\ee
while the asymptotic decay is described by 
\be\label{tail}
f_s(x)\sim |x|^{-3/2}\exp(-|x|/4l_{\infty});\,\,\, |x| \gg 4l_{\infty}, 
\ee
revealing that the decay rate $s(x) = d \log f_s/dx$ changes by a factor~$4$.

\begin{figure}
\begin{center}
\epsfig{figure=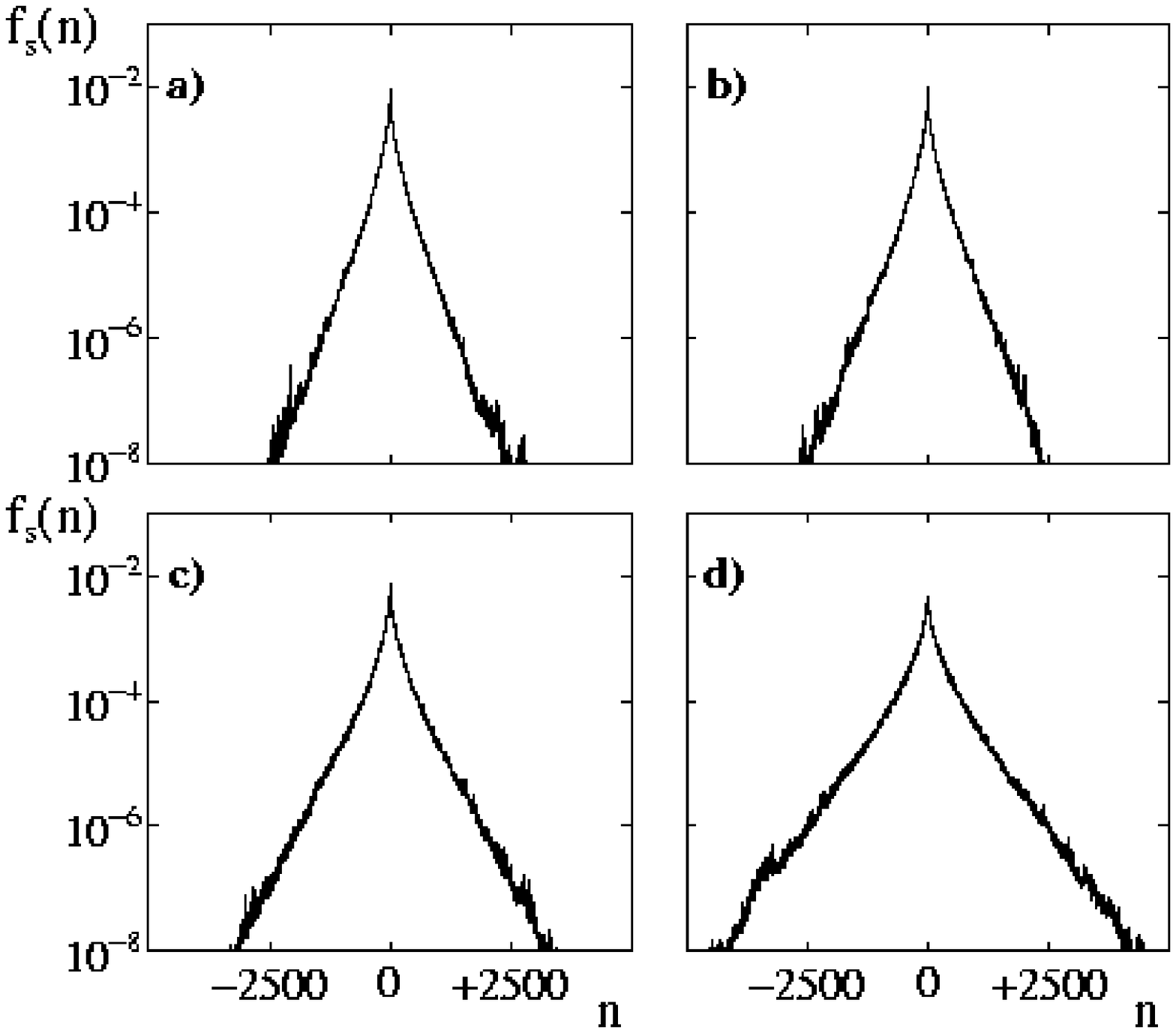,width=8cm}\hfill
\end{center}
\noindent
{\footnotesize {\bf FIG.~1.}
Arithmetic average of the asymptotic wavefunction~$f_s(n)$. The bandwidth is $b=15$,
and the snapshot is taken at $t=10^4$. The data correspond to various
crossover parameters: (a) $\alpha=0$; (b) $\alpha=0.02$; (c) $\alpha=0.085$; 
and (d) $\alpha=0.7$.}
\end{figure}

In Fig.~1 we report some asymptotic profiles~$f_s(x,\alpha)$ for~$b=15$ and various 
values of the time-reversal symmetry breaking parameter~$\alpha$. These data strongly 
suggest that the decay of~$f_s(n)$ in the vicinity of the origin is definitely faster 
than in the tails. As no analytical results are available for the crossover
regime, it is very tempting to compare our numerical results with the theoretical 
dependence~(\ref{Gogolin}) derived for quasi-1D systems with pure symmetries. 
The profiles reported in Fig.~2a are the result of an arithmetic average over 
many realizations of the asymptotic profile for $b=5,10,15$ and several~$\alpha$. 
They are plotted with the scaling assumption
\be\label{prof}
f_s(n,t\rightarrow \infty )=l_{\infty}f_s(x);\quad x\equiv n/l_{\infty} 
\ee
where~$l_\infty$ is determined by a fit according to Eq.~(\ref{Gogolin}).
The very good agreement between the numerical results and the analytical curve over a 
broad range of $x$-values suggests that a properly modified theory including the 
effect of an intermediate magnetic field should be able to account for the asymptotic
profile of wavepackets in quasi-1D systems. Furthermore the scaling relation~(\ref{prof}) 
implies that the wavefunction in the crossover regime shows the same gross structure 
(envelope) as in the pure symmetry cases, on scales comparable to the 
localization length. This is in contrast to the occurrence of a second localization 
scale for the far tails proposed in~\cite{KE99,KE00}. Hence we expect that all the scaling 
laws for the eigenstates dominated by the fluctuations of the "envelope" (e.g., moments 
like the inverse participation ratio), also hold for the crossover regime.

\begin{figure}
\begin{center}
\epsfig{figure=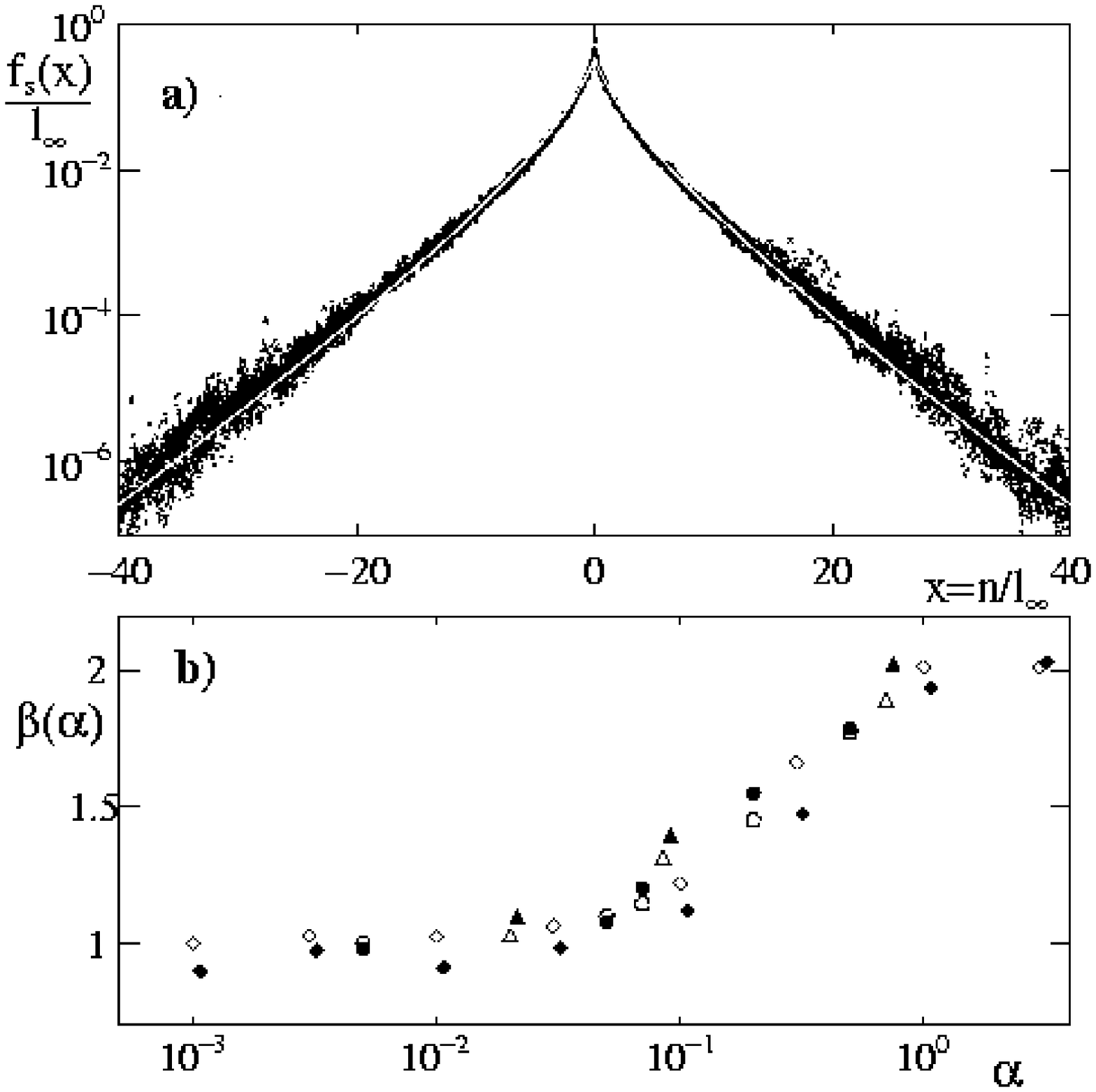,width=8cm}\hfill
\end{center}
\noindent
{\footnotesize {\bf FIG.~2.}
(a) Asymptotic average profile of the wavepackets for $b=5,10,15$ and various
crossover parameters~$\alpha$ rescaled according to Eq.~(\ref{prof}). The light 
smooth line is the theoretical expression~(\ref{Gogolin}). (b) The localization 
lengths~$l_{\infty}$ (open symbols) and $l_{\infty}^{\rm as}$ (full symbols) as 
a function of the time reversal symmetry breaking parameter~$\alpha$. Various
symbols correspond to various values of~$b$; diamonds to~$b=5$, circles to~$b=10$
and triangles to~$b=15$.
} 
\end{figure}

In Fig.~2b the asymptotic localization length~$l_{\infty}$, as determined from 
the best fit with~(\ref{Gogolin}) is plotted versus the crossover parameter~$\alpha$. 
One can clearly see that the transition from preserved to broken time-reversal 
symmetry is rather smooth and thus we have
\be\label{lalpha}
l_{\infty}(\alpha) = \beta(\alpha) l_{\infty}(0)
\ee
where~$\beta(\alpha)$ is a smooth function that interpolates between the values~$1$ 
and~$2$ for preserved ($\alpha=0$) and totally broken time reversal symmetry 
($\alpha\simeq 1$) respectively. In Fig.~2b we also present the localization 
length derived by a direct fit of the average of the logarithm of the asymptotic 
wavefunction $l_{\infty}^{\rm as}= \lim_{N\rightarrow \infty} \langle\ln(f_s)
\rangle/(2N)$. It clearly shows the same smooth behaviour as $l_{\infty}$. 

Our numerical results are in contrast to the supersymmetry results~\cite{KE99,KE00} 
and agree nicely with the transfer matrix calculations~\cite{SB99}. Moreover, they
give a definite answer to the question, whether the assumed discrepancy is due 
to the fact that in~\cite{KE99,KE00} the logarithm of the averaged wavefunction was 
investigated, while in~\cite{SB99} it was the logarithmic average. Based 
on our calculations we are now able to exclude both possibilities suggested 
in~\cite{KE00,SB99}; namely that one could attribute the two-scale localization phenomenon 
to anomalously localized states, that dominate the asymptotic average profile~$f_s(n)$ or 
to the non-applicability of the Borland conjecture in the crossover regime.

\begin{figure}
\begin{center}
\epsfig{figure=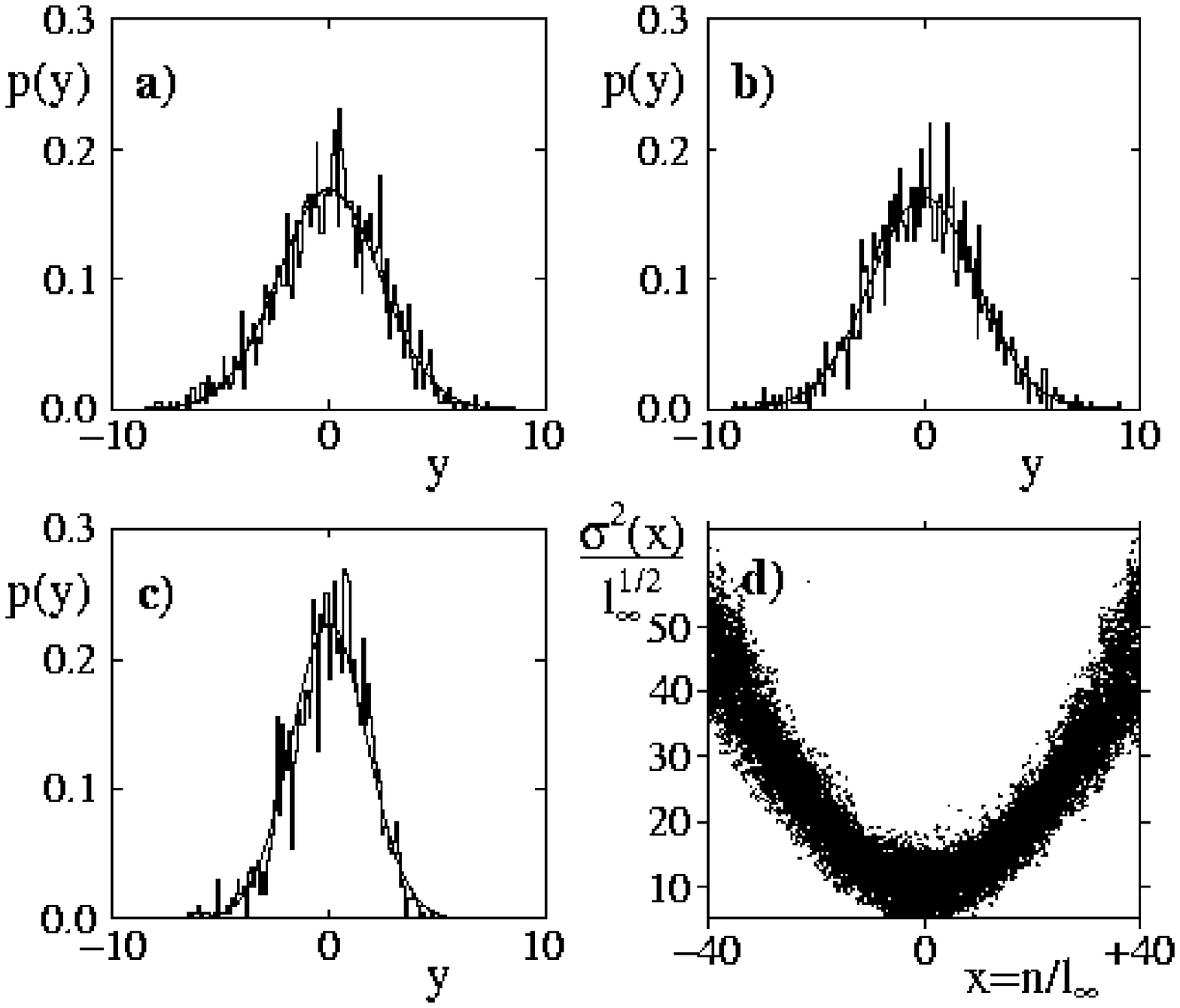,width=8cm}\hfill
\end{center}
\noindent
{\footnotesize {\bf FIG.~3.}
Distribution of $y\equiv \ln f_s(n=6b^2\pm 2)$ for~$b=5$ and various values for
the time-reversal symmetry breaking parameter: (a) $\alpha=0.003$; (b) $\alpha=0.03$; 
(c) $\alpha=0.3$; (d) Variance of~$y$ for $b=5,10,15$ and various $\alpha$. The data are rescaled 
according to Eq.~(\ref{varsc}).}
\end{figure}

As mentioned above, the localization of all eigenstates implies that for $t\rightarrow 
\infty$ the quantum steady state~$f_s(n)$ is localized and fluctuates around the average 
profile~(\ref{Gogolin}). Thus, one can ask about the distribution of~$f_s(n)$. Having in 
mind that the asymptotic profile is exponentially localized (see Eq.~(\ref{Gogolin})), 
we computed the distribution of the logarithm of the wavefunction $y\equiv \ln(f_s)$ for 
various values of the crossover parameter~$\alpha$. In Figs.~3a-c we report some 
representative distributions ${\cal P} (y)$, which refer to the case~$b=5$ with 
$\alpha= 0.003,0.03,$ and $\alpha =0.3$. In all cases, we have found that the distribution of the
logarithm of the wavefunction ${\cal P} (y)$, is a Gaussian with good accuracy. This is 
in perfect agreement with the analytical calculation of~\cite{KE00} for ${\cal P} (y)$.
In particular a useful indicator is, the spatial growth of the asymptotic variance
\be\label{varexp}
\sigma ^2(n)=\langle(\ln f_s(n))^2\rangle-\langle\ln f_s(n)\rangle^2  \;.
\ee
The results for the cases~$b=5,10$ and various values~$0\le\alpha\le1$ are reported in Fig.~3d 
under the scaling assumption 
\be\label{varsc}
\sigma^2(n) = \sigma^2(x)/\sqrt{l_\infty};\,\, n/l_{\infty}.
\ee
This further confirms that not only the mean asymptotic profile~$f_s(n)$ but also 
higher moments change smoothly as a function of the crossover parameter~$\alpha$.

A global characterization of the diffusion and localization properties of a wavepacket is 
provided by the evolution of the mean-square displacement 
\be\label{m2}
M(t,\alpha)\equiv \left\langle \sum_mm^2|c_m(t)|^2\right\rangle \quad , 
\ee
where~$\langle\cdot\rangle $ denotes the average over different realizations of the 
disorder. On the basis of the localization properties of the asymptotic wavefunction 
(see Eqs.~(\ref{Gogolin},\ref{lalpha})) one expects that, for~$b\gg 1$, $M_{\infty}$~grows as 
$M_{\infty}\sim l_{\infty}(\alpha)^2$. The quantity~$M$ reaches its maximum value~$M_\infty$
at $t\simeq t_D$. Up to that time the evolution of the packet is diffusive, i.e.
$M_\infty \simeq  M(t_D) \simeq D t_D$, where~$D$ is the diffusion constant. From
supersymmetry~\cite{E97} it is known that $l_{\infty} \Delta = D$ with~$\Delta$ being the mean
level spacing. The latter depends on~$\alpha$ as $\Delta(\alpha) = {\sqrt {1+0.5
\alpha^2}} \Delta(0)$. Accordingly the following relations hold:
\be\label{tdD}
\frac{t_D(\alpha)}{t_D(0)}= \frac {\beta(\alpha)}{\sqrt{1+0.5\alpha^2}} \,\,;\,\,\,
\frac{D(\alpha)}{D(0)}=\beta(\alpha){\sqrt{1+0.5\alpha^2}} 
\ee
where $t_D(0)\sim b^{3/2}$ and $D(0)\sim b^{5/2}$~\cite{IKPRT96}. In the inset of Fig.~4 
we show the scaling of the diffusion constant~$D(\alpha)$ for $b=30,40$ and various~$\alpha$. 
Our numerical results are in perfect agreement with~(\ref{tdD}).
Equation~(\ref{tdD}) suggests that the mean-square displacement~$M(t)$ follows the scaling 
relation
\be\label{variance}
M(t,\alpha) = M_\infty(\alpha) {\tilde M}(t/t_D(\alpha)).
\ee
The numerical results obtained for the cases $b=5,10,15$ and various values of the 
crossover parameter~$\alpha$ are reported in Fig.~4 according to the above ansatz. The 
close coincidence of data reveals the presence of a scaling regime already at 
moderately large $b$-values. 

\begin{figure}
\begin{center}
\epsfig{figure=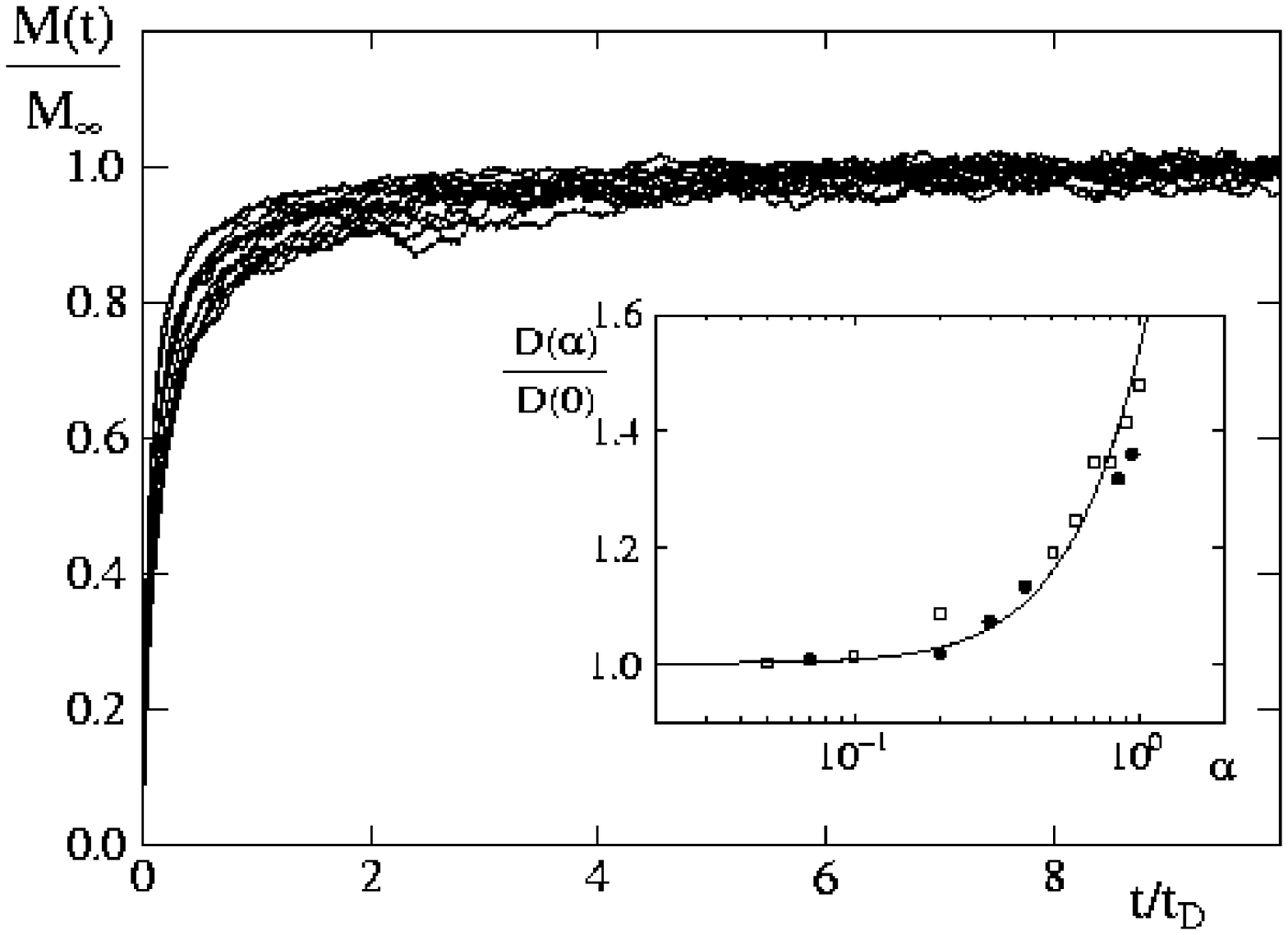,width=8cm}\hfill
\end{center}
\noindent
{\footnotesize {\bf FIG.~4.}
Mean-square displacement~$M(t)$ for $b=5,10,15$ and various values of~$\alpha$,
confirming the scaling law~(\ref{variance}). 
In the inset we present the diffusion constant~$D(\alpha)$ as a function of the time-reversal 
symmetry breaking parameter~$\alpha$ for $b=30$ (squares), $40$ (filled circles). The 
solid line is the theoretical expectation~(\ref{tdD}).}
\end{figure}

In conclusion, we have studied scaling properties of both diffusion and strong localization 
of wavepackets in quasi-1D (i.e., in 1D random media with long-range interactions). Our 
numerical data show that the decay rate of the asymptotic wavefunction profile changes 
smoothly, in contradiction to the supersymetry results~\cite{KE99,KE00}. We have also found 
that the asymptotic shape of the wavepacket is well reproduced by the analytical 
expression~(\ref{Gogolin}) derived for the pure symmetry cases (i.e.~$\alpha=0,1$). Moreover, 
in agreement with~\cite{KE00} we find, that the logarithm of the asymptotic wavefunction is 
normally distributed for every~$\alpha$, and the variance is scaled according to Eq.~(\ref{varsc}). 
Another issue addressed in this Letter concerns the wavepacket evolution. In particular, 
we found how the diffusion constant~$D(\alpha)$ and the diffusion time~$t_D(\alpha)$ change 
as a function of the time reversal symmetry breaking parameter~$\alpha$. As a result we 
were able to establish a scaling law~(\ref{variance}) for the variance~$M(t)$ of the packet.

We are grateful to Y. Fyodorov and A. Kolesnikov for useful discussions and suggestions.


\end{multicols}
\end{document}